\documentclass[final,5p,times,twocolumn]{elsarticle}

\usepackage{amssymb}
\usepackage{amsthm}
\usepackage{tikz}
\usepackage{bm}
\usepackage{pgfplotstable}
\usepackage{pgfplots}
\usepgfplotslibrary{groupplots}

\usepackage{color}
\usepackage{lineno}

\journal{Powder Technology}

\begin{document}

\begin{frontmatter}



\title{Particle dynamics simulation of wet granulation in a rotating drum}


\author[a,b]{Thanh Trung Vo}
\author[a]{Saeid Nezamabadi\corref{cor1}}
\cortext[cor1]{Corresponding author:}
\ead{saeid.nezamabadi@umontpellier.fr}
\author[c]{Patrick Mutabaruka}
\author[d]{Jean-Yves Delenne}
\author[e]{Edouard Izard}
\author[c]{Roland Pellenq}
\author[a]{Farhang Radjai}

\address[a]{LMGC, Universit\'e de Montpellier, CNRS, Montpellier, France.}
\address[b]{Bridge and Road Department, Danang Architecture University, 553000 Da Nang, Vietnam.}
\address[c]{$\langle \mbox{MSE} \rangle^2$, UMI 3466 CNRS-MIT, MIT Energy Initiative, 77 Massachusetts Avenue, Cambridge 02139, USA.}
\address[d]{IATE, UMR1208 INRA - CIRAD - Universit\'e de Montpellier - SupAgro, 34060 Montpellier, France.}
\address[e]{ArcelorMittal R\&D Maizi\`eres, Voie Romaine, F-57283, Maizi\`eres-L\`es-Metz, France.}

\begin{abstract}
We simulate the granulation process of solid spherical particles in the presence of a viscous liquid 
in a horizontal rotating drum by using molecular dynamics simulations in three dimensions. 
The numerical approach accounts for the cohesive and viscous effects of the  binding liquid, 
which is assumed to be transported by wet particles 
and re-distributed homogeneously between wet particles in contact. We investigate the growth of 
a single granule introduced into the granular bed and the cumulative numbers of accreted 
and eroded particles as a function of time for a range of values of material parameters such 
as mean particle size, size polydispersity, friction coefficient and liquid viscosity. 
We find that the granule growth is an exponential function of time, reflecting 
the decrease of the number of free wet particles. The influence of material parameters on the 
accretion and erosion rates reveals the nontrivial dynamics of the granulation process. It opens the 
way to a granulation model based on realistic determination of particle-scale mechanisms of granulation.
\end{abstract}


\end{frontmatter}

\section{Introduction}
\label{intro}

Wet granulation or agglomeration of fine solid particles into larger particles, 
called granules or agglomerates, is a widespread technique in 
industrial processes such as the manufacture of 
pharmaceuticals \cite{Chien2003}, fertilisers and food products \cite{Rondet2012,Rondet2010}, 
powder metallurgy \cite{Nosrati2012} and 
iron-making \cite{SASTRY2003,Aguado2013,WALKER2007}. 
The increased size of the granules modifies the rheological properties of 
the granular material and may improve 
flow properties, reduce the segregation of different types of primary 
particles or enhance the permeability for the interstitial gas between grains 
\cite{Rondet2010,SalehGuigon2007,Suresh2017,Iveson2001,Herminghaus2005,Litster2014,Liao2016}. 
The wetting of primary particles is achieved either by mixing them with a binding liquid prior to 
the process or by dripping or spraying the liquid to the material during the process  \cite{Liao2016,Iveson1998,Xiao2017,Basel2016,Bouwman2005b,Degreve2006,Osborne2011,Pawar2016,Pashminehazar2016}. 
The granules nucleate as a result of the collisional-frictional/capillary-viscous interactions of 
wet primary particles and increase in size by incorporation of the available liquid, accretion of 
primary finer particles and coalescence with other granules \cite{Ennis1991}. 
The existing granules may disappear or keep their size depending on the amount of 
available liquid and the rate of erosion as compared to that of 
accretion and coalescence \cite{Stepanek2008b}.  

Given the large number of parameters involved in the granulation process, 
its detailed physical mechanisms and their relative importance for the resulting properties of 
are complex. One may distinguish two different groups of parameters: process and material  \cite{SalehGuigon2007,Suresh2017,Ghadiri2011a,Ennis1996,Butensky1971}. 
The material parameters are the properties of the binding liquid and raw material  
such as liquid viscosity, primary particle size distribution, mean particle size and friction coefficient of 
primary particles \cite{Osborne2011,Ileleji2016,Stepanek2008a}. The process parameters 
are related to the method of mixing solid particles with the binding liquid and the 
corresponding operating parameters. There are different types of granulators 
using fluidized beds, high shear or low shear in planetary devices or rotating drums or disks \cite{SalehGuigon2007}. 
Major process parameters in all types of granulators are the liquid volume \cite{WALKER2007,Ghadiri2011a,Ileleji2016}, 
granulator size, rotation speed, inclination angle \cite{Pan2006,Spurling2001,Wang2007a} and filling rate \cite{Gray2001}.  
The granulation process often needs to be optimized by playing with all these parameters 
in order to produce granules of high density, homogeneous distribution 
of primary raw particles, a targeted mean size and high strength 
\cite{Thornton1998,Liu2000j,Iveson2002,Ghadiri2011,Rojek2017}. 

The granular dynamics has not the same characteristics in different granulator geometries  
and may favor more or less collisional or frictional contacts between primary particles 
and influence the redistribution and transport of the binding liquid. The granulation 
process is easier to model and control when the agglomeration is governed by 
binary collisions between particles, as in granulators based on fluidized bed or 
high shearing by impellers. Such processes have been extensively investigated in 
application to the pharmaceutical industry \cite{Suresh2017}. In contrast, 
in drum granulators the particles agglomerate in a downward dense granular flow 
along inclined rotating drum. The drum agglomeration has the advantage of 
being a continuous and robust process, but since the rheology of dense granular flows is 
a matter of current research \cite{GDR_MiDi_2004,Kamrin2015,Amarsid2017}, 
the agglomeration mechanisms in this geometry remain quite poorly understood \cite{SalehGuigon2007}.   
Granular flows in an inclined rotating drum may show several flow regimes \cite{Govender2016,Yang2008,Mellmann2001}  
with the common feature of being dense and inhomogeneous, and involving inertial effects  \cite{Iveson1998,Degreve2006,Stepanek2008b,Lian1993,Scheel2008,Willett2000,Delenne2012}.
A practical difficulty with drum granulation is the in-line monitoring of the kinetics, 
making it less amenable to theoretical understanding, which is required 
in order to be able to improve drum granulation plants, often suffering from  
a significant recycle of undersize and crushed oversize granules \cite{SalehGuigon2007}. 

In this paper, we present extensive simulations of drum granulation by means of the  
District Element Method (DEM) \cite{Radjaibook}. The use of the DEM, in which the granular material 
is modeled as an assembly of spherical particles interacting via frictional/cohesive 
forces, allows for direct quantification of particle-scale kinetics and 
accretion/erosion events. This method has already been applied to 
investigate agglomeration in granular shear flows. For example, Talu et al. \cite{Talu2000} 
introduced a model of wet granulation in which some of the particles which are assumed 
to be covered by a binder and therefore sticky while the rest are dry. 
The binder-layer between particles dissipates energy due to viscosity and  
and allows the particles to stick to another one by the action of capillary forces. 
In other reported DEM simulations of the granulation process, besides capillary and viscous forces, 
simple empirical rules are used for progressive wetting of the particles \cite{Chan2016,LauKind2016,Sarkar2018}. 
To reduce the high computational cost of DEM simulations, some authors have 
used the DEM simulation data with a low number of particles to train an artificial 
neural network or in conjunction with population balance equations for application to 
the large number of particles \cite{LauKind2016,Barrasso2014}. 
The DEM has, however, never been employed for drum granulation. 

In the following, in section \ref{sec:method_model}, we introduce our 
numerical approach and a model for the 
transport and redistribution of the binding liquid. We analyze in section \ref{sec:behavior} 
the evolution of granules as function of the number of drum rotations by investigating 
the effects of the process and material parameters such as rotation speed, Froude number, 
size ratio between large and small particles, mean particle size, friction coefficient and  
viscosity of the binding liquid, we characterize the 
dynamics of granulation in terms of  the rate of accretion and erosion. 
Finally, we conclude in section \ref{sec:conclude} with a short summary 
and possible further research directions.

\section{Model description and numerical method}
\label{sec:method_model}

Our numerical approach for the simulation of the agglomeration process 
in a rotating drum is based on the molecular method and a model for 
the redistribution and transport of the binding liquid. We first describe below 
the physical assumptions underlying the model. Then, we briefly present the 
numerical algorithm with its input parameters and main calculation steps.     

\subsection{Physical assumptions}
\label{subsec:assumptions}

In the molecular dynamics (MD) method, the equations of motion of all particles 
are integrated according to an explicit time-stepping scheme such as 
the well-known velocity-Verlet algorithm 
\cite{Radjaibook,Richefeu2009a,Matuttis2000,Thornton2000a,Radjai2001a,Herrmann1998a}. 
A detailed description of the liquid phase and its interaction with solid particles, 
requires sub-particle discretization of the liquid phase and a numerical model for 
liquid-gas phase transition \cite{Scheel2008,Delenne2012,Amarsid2017}. 
However, such a multi-component model of partially saturated granular materials 
is not computationally efficient for the simulation of the granulation process 
involving a large number of primary particles. An efficient alternative approach 
consists in accounting for the capillary and lubrication forces between particles 
as a well as a particle-scale model for the distribution and transport of the liquid. 

Recent experiments and numerical simulations show that the liquid 
clusters condensed from a vapor or introduced by mixing the liquid with grains    
can be characterized by their connectivity with the grains. The number of 
liquid clusters connected to two grains prevails for low amounts of the liquid. 
In this `pendular' state, the liquid is in the form of binary bridges. 
As the amount of liquid increases, the clusters involve more and more 
particles until a single cluster spans the whole packing. The cohesive effect of 
the liquid in thermodynamic equilibrium is controlled by the total wetted 
surface and the Laplace pressure. The cohesion rapidly increases as the amount of liquid 
is increased in the pendular state, and then it keeps a nearly 
constant value (or slightly increases) with increasing amount of 
the liquid before declining for large amounts of the liquid \cite{Scheel2008,Delenne2012}. 
This description assumes, however, that the particles are in 
quasi-static equilibrium and the liquid is in thermodynamic equilibrium. 
The negative Laplace pressure within the liquid phase is not uniformly 
distributed if the system is out of equilibrium. Furthermore, if the granular material 
flows (as inside a rotating drum), the liquid clusters undergo large distortions,   
and the liquid is continuously re-distributed as a result of coalescence and separation of 
liquid clusters. In practice, a small amount of the added liquid is  
adsorbed into the particle rough surfaces and is not directly 
involved in capillary bonding between particles. 

These features suggest that  in a dense granular material, for a broad range of the amounts of liquid, 
the cohesive capillary stress is nearly independent of the amount of liquid, and therefore the 
effect of liquid volume can be accounted for by the number of wet particles. 
In a rotating drum, when the liquid is poured onto the granular flow, it has not time to diffuse and   
the capillary stress leads to the creation of small aggregates of primary 
particles that are transported by the granular flow. These ``micro-aggregates" may deform 
or break up into smaller aggregates. They may also capture more primary particles or coalesce 
into larger aggregates if they have an excess amount of liquid that can be shared 
with other micro-aggregates as a consequence of their consolidation under the 
action of contact forces inside the granular flow. The initial size of the micro-aggregates is proportional to that 
of the droplets but, depending on the wetting method, it can grow rapidly into granules. 

\begin{figure}
\centering
\includegraphics[width=0.47\textwidth,clip]{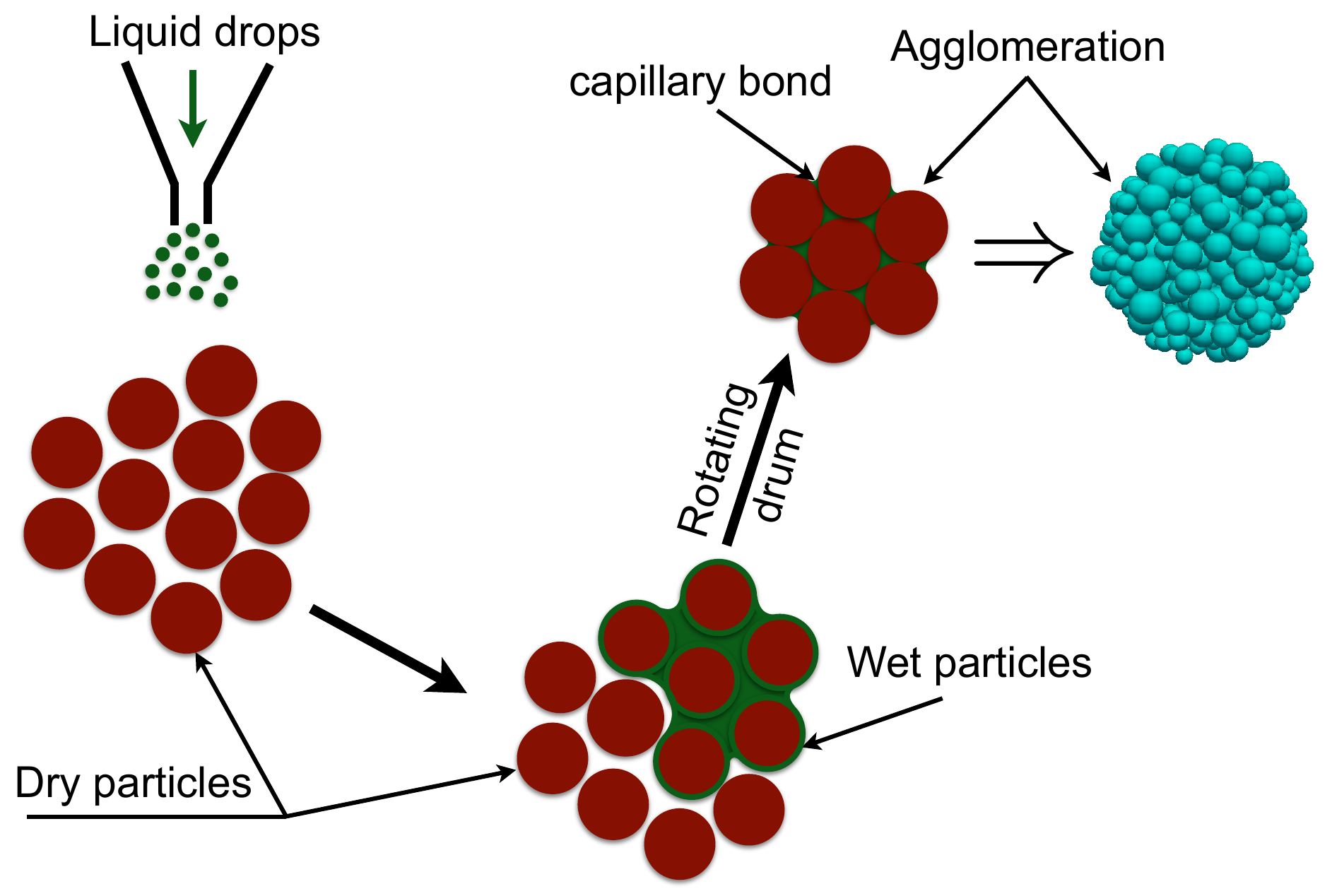}
\caption{Schematic representation of the granulation model.}
\label{fig:agglomeration_process}
\end{figure}

This picture of liquid transport by micro-aggregates, illustrated in Fig. \ref{fig:agglomeration_process}, 
means that the DEM simulations can be based on the micro-aggregates as wet units.  
In its simplest setting, these basic units can be modeled as particles each transporting 
a given amount of liquid. Since the number of primary particles embodied 
in each micro-aggregate is nearly proportional to the droplet size, the size distribution of 
the basic particles can be regarded as reflecting that of the droplets. 
When two wet particles (micro-aggregates) meet, they are subjected to the cohesive action of 
the spontaneous liquid bridge appearing between them. In this sense, the interactions between 
micro-aggregates are similar to those between particles covered by a liquid layer as in 
the method introduced by Talu et al. \cite{Talu2000}. In the following, we present in more detail 
our numerical method and all model parameters.            

\subsection{Numerical method}
\label{subsec:method}

In the MD method, the particles are modeled as rigid grains interacting via 
visco-elastic forces reflecting the contact force and the local contact strain defined from 
the relative displacement of particles. Since the particles are assumed to be rigid, 
a large repulsive stiffness and hence a high time resolution are required for the calculation of 
the interactions between particles. The motion of each rigid spherical particle $i$ of 
radius $R_i$ is governed by Newton's second law under the action of normal contact forces $f_n$, 
tangential contact forces $f_t$, capillary forces $f_c$, viscous forces $f_{vis}$ and particle weight $m_i\mathbf{g}$ : 
\begin{eqnarray}
m_i \frac{d^2 \mathbf{r}_i}{dt^2} &=&  \sum_j [(f_n^{ij} + f_c^{ij} + f_{vis}^{ij})\mathbf{n^{ij}} + 
f_t^{ij} \mathbf{t}^{ij}] + m_i\mathbf{g} \nonumber \\
\mathbf{I}_i \frac{d \bm \omega_i}{d t}  &=& \sum_j f_t^{ij} \mathbf{c}^{ij} \times \mathbf{t}^{ij}
\label{eq:motion}
\end{eqnarray}
where $\bm \omega_i$ is the rotation vector of particle $i$, and 
$m_i$, $\mathbf{I}_i$, $\mathbf{r}_i$ and $\mathbf{g}$ are the mass, inertia matrix, 
position and gravity acceleration vector of particle $i$, respectively. $\mathbf{n}^{ij}$ denotes 
the unit vector perpendicular to the contact plane with particle $j$ and pointing from $j$ to $i$,  
$\mathbf{t}^{ij}$ is the unit vector belonging to the contact plane $ij$ and pointing in the direction 
opposite to the relative displacement of the two particles and $\mathbf{c}^{ij}$ is the vector 
pointing from the center of particle $i$ to the contact point with particle $j$. 
The tangential viscous dissipation, as compared to the normal lubrication force, 
is neglected  \cite{lefebvre2013}. The equations of motion are integrated according to 
a velocity-Verlet time-stepping scheme \cite{Radjaibook,duran1999sands}.

\begin{figure}[tbh]
\centering
\includegraphics[width=6.5 cm]{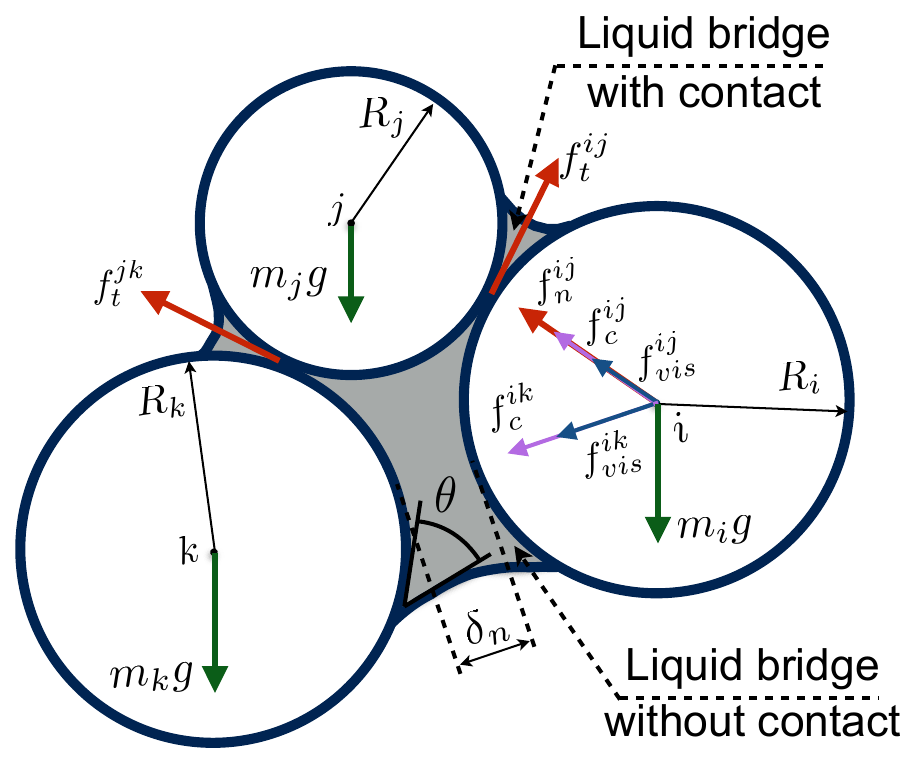}(a)
\includegraphics[width=8.2 cm]{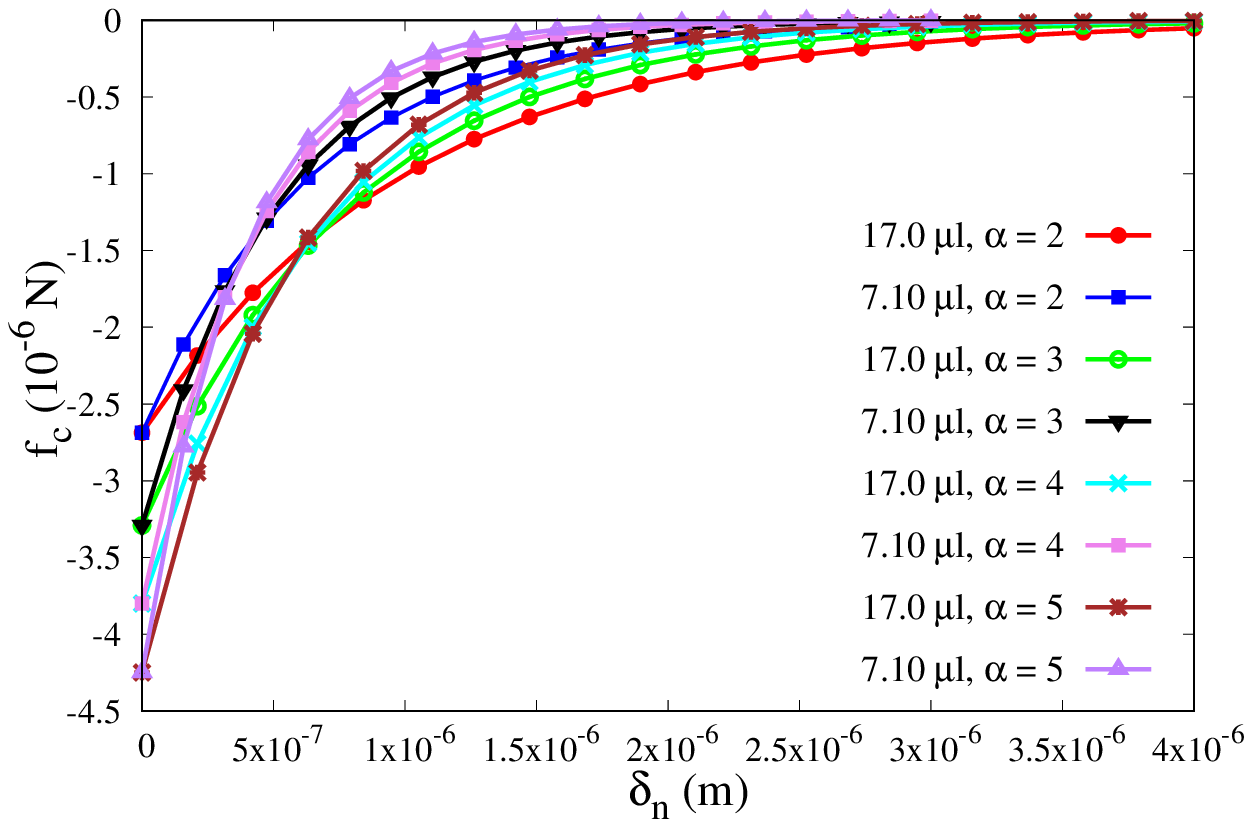}(b)
\caption{(a) Schematic drawing of two different cases of capillary bridges: particle $i$ in 
contact with particle $j$ and without contact with particle $k$; 
(b) Capillary cohesion force $f_c$ as a function of the gap $\delta_n$ between two 
particles for different values of the liquid volume $V_b$ ($\mu$l) and size ratio $\alpha$.}
\label{fig:geometry}
\end{figure}

The normal contact force $f_n$ involves two components \cite{Richefeu2006,Trung2017a}:
\begin{equation}
\displaystyle{f_n = f_n^e + f_n^d} \ .
\label{eq:fn}
\end{equation}
The normal elastic force $f_n^e=k_n \delta_n$ is a linear function of the normal elastic deflection 
$\delta_n$, where $k_n$ is the normal stiffness constant, and the normal damping force $f_n^d= \gamma_n \dot{\delta}_n$ 
is proportional to the relative normal velocity $\dot{\delta}_n$, where $\gamma_n$ is the normal 
viscous damping parameter. These both forces appear only when there is overlap, i.e. for $\delta_n < 0$.

As the normal force, the tangential contact force $f_t$  is the sum of an elastic force $f_t^e = k_t \delta_t $ 
and a damping force $f_t^d = \gamma_t \dot{\delta}_t $, where $k_t$ is the tangential 
stiffness constant, $\gamma_t$ denotes the tangential viscous damping parameter and $\delta_t$ 
and $\dot{\delta}_t$ are the tangential displacement and velocity in contact, respectively. 
According to the Coulomb friction law, the contact tangential force is bounded by  
a force threshold $\mu f_n$, where $\mu$ is the friction coefficient\cite{Richefeu2007a,Schaefer1996,Dippel1997b,Luding1998c}:
\begin{equation}
f_t = -\!\min \left\{ (k_t\delta_t + \gamma_t \dot{\delta}_t), \mu f_n \right\}.
\label{eq:ft}
\end{equation}

The capillary attraction force $f_c$ between two particles depends on the liquid volume $V_b$ of the liquid bond, 
liquid-vapor surface tension $\gamma_s$ and particle-liquid-gas contact angle $\theta$; see Fig. \ref{fig:geometry} 
\cite{Lian1993,Richefeu2006,Richefeu2007b}. The capillary force is computed from the 
Laplace-Young equations. In the pendular state, an approximate solution is given by 
the expression \cite{Richefeu2007a,Trung2018a} :
\begin{equation}
	f_c = \left\{
	\begin{array}{@{}ll@{}}
	-\kappa\  R,						&  \mbox{for}\  \delta_n <  0, \\
	-\kappa\  R\  e^{-\delta_n /\lambda},  	&  \mbox{for}\  0  \leq  \delta_n  \leq  d_{rupt}, \\
	0,	 							&  \mbox{for}\  \delta_n > d_{rupt},
	\end{array}\right.
\label{eq:fc}
\end{equation}
where $R = \sqrt{R_i R_j}$ is the geometrical mean radius of two particles of radii $R_i$ and $R_j$ 
and the capillary force pre-factor $\kappa$ is \cite{Richefeu2007a}:
\begin{equation} 
\kappa = 2\pi \gamma_s \cos \theta.
\label{eq:kappa}
\end{equation}
This force exists up to a debonding distance $d_{rupt}$ given by \cite{Richefeu2006,Trung2017a}  
\begin{equation}
\displaystyle{d_{rupt} = \left(1+\frac{\theta}{2}\right)V_b^{1/3} } \ .
\label{eq:d_rupt}
\end{equation}
The characteristic length $\lambda$ is the factor that presents the exponential falloff of the 
capillary force in equation~(\ref{eq:fc}):
\begin{equation}
\lambda = c\ h(r)\ \left(\frac{V_b}{R'}\right)^{1/2}
\label{eq:lamda}
\end{equation}
Here, $R'= 2R_iR_j/(R_i+R_j)$ and $r\!=\!\max \{ R_i/R_j ; R_j/R_i\}$ are the harmonic 
mean radius and the size ratio between two particles. 
The expression~(\ref{eq:fc}) nicely fits the capillary force as obtained from direct integration of the 
Laplace-Young equation by setting $h(r) = r^{-1/2}$ and $c \simeq 0.9$ \cite{Richefeu2006,Richefeu2007a,Radjai2009}. 
During the simulation, the total amount of liquid is evenly re-distributed among all pairs of 
wet particles (or micro-aggregates) having a gap below $d_{rupt}$. The largest value of 
the capillary force occurs when two particles are in contact ($\delta_n \leq 0$). 
This is the case of most liquid bonds, and the capillary force at those 
contacts is independent of the liquid volume \cite{Radjai2009}.

The normal viscous force $f_{vis}$ is due to the lubrication effect of liquid bridges between particles.  
Its classical expression for two smooth spherical particles is \cite{lefebvre2013,HappelBrenner1983}:
\begin{equation}  
f_{vis} = \frac{3}{2} \pi R^2 \eta \frac{{v_n}}{\delta_n},
\label{eq:f_vis}
\end{equation}
where $\eta$ is the liquid viscosity and $v_n$ is the relative normal velocity  assumed to be 
positive when the gap $\delta_n$ is decreasing. This expression implies that the viscous force diverges  
when the gap $\delta_n$ tends to zero. With this singularity, two rigid particles can not collide  in  
finite time. However, for slightly rough particles, the surfaces are no longer parallel and the characteristic 
size of the asperities allows for collision in finite time. Hence, we introduce 
a characteristic length $\delta_{n0}$ representing the size of asperities and assume that the 
lubrication force is given by 
\begin{equation}  
f_{vis} = \frac{3}{2} \pi R^2 \eta \frac{{v_n}}{\delta_n+\delta_{n0}} \; \; \; \;  \mbox{for} \; \; \delta_n > 0
\label{eq:f_vis2}
\end{equation}   
as long as $\delta_n >0$, i.e. for a positive gap. This expression ensures that the singularity 
will not occur as long as there is no contact. When contact occurs, i.e. for $\delta_n <0$, we assume that 
the lubrication force depends only on the characteristic length, so that  
\begin{equation}  
f_{vis} = \frac{3}{2} \pi R^2 \eta \frac{{v_n}}{\delta_{n0}} \; \; \; \; \mbox{for} \; \; \delta_n \leq 0.
\label{eq:f_vis3}
\end{equation}     
In our simulations, we set $\delta_{n0} = 5.10^{-4} d_{min}$. This value is small enough 
to allow lubrication forces to be effective without leading to its divergence 
at contact.

\subsection{Granulation process model}
\label{subsec:model}

\begin{figure}[tbh]
\centering
\includegraphics[width=0.42\textwidth,clip]{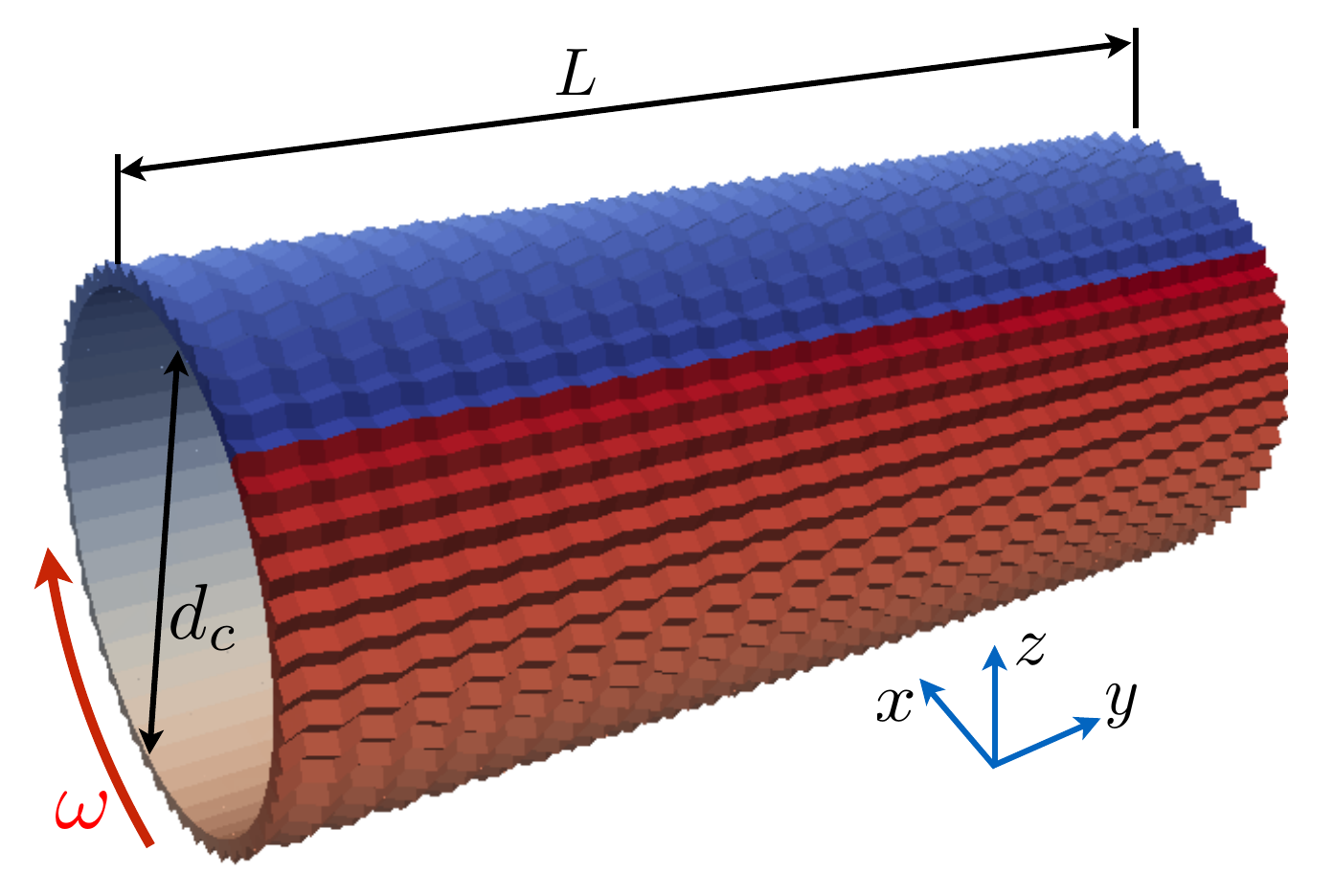}
\caption{The geometry of the numerical model of a horizontal rotating 
drum used in the granulation simulations.}
\label{fig:drum}
\end{figure}

The rotating drum is a cylinder of length $L$ and diameter $d_c$ constructed geometrically 
by the juxtaposition of polyhedral rigid elements. Its both ends are closed by two planes and 
it can rotate around its axis at given angular speed $\omega$; see Fig. \ref{fig:drum}. 
In all simulations analyzed below, the drum is horizontal, implying that the gravity is perpendicular 
to the drum axis. 

\begin{figure}[tbh]
\centering
\includegraphics[width=0.35\textwidth,clip]{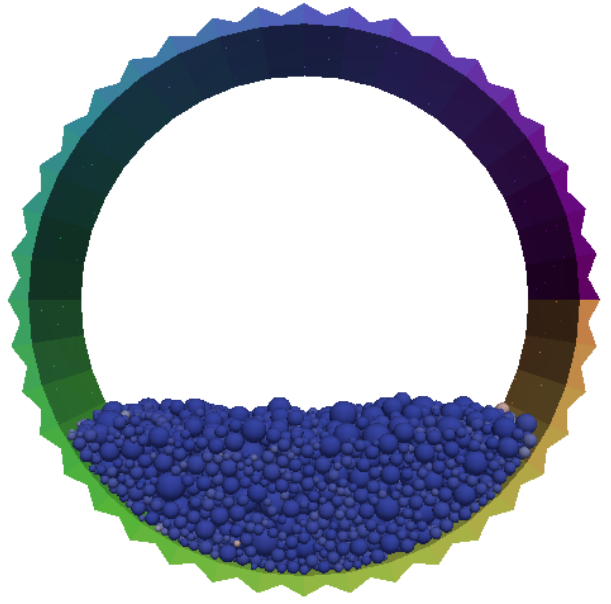}(a)
\includegraphics[width=0.35\textwidth,clip]{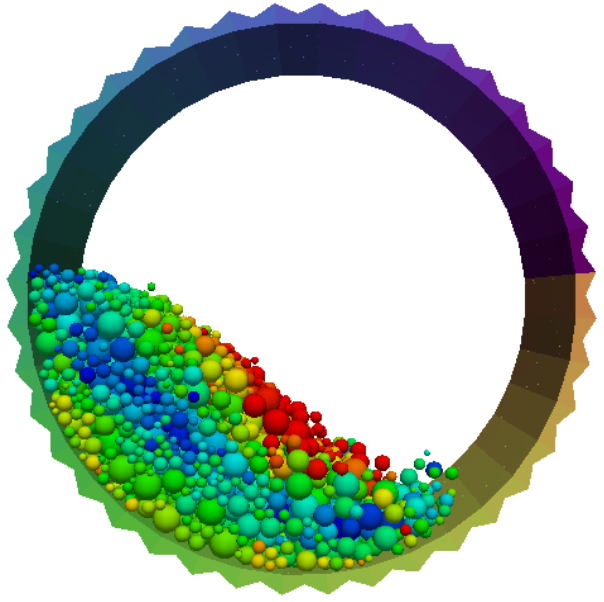}(b)
\caption{Snapshots of the initial state (a) and stable flow (b) in the drum for 
$\alpha=5$. The colors show the magnitudes of particle velocities varying from 
red (fast particles at the free surface) to blue (slowest particles in the middle of the drum).}
\label{fig:process}
\end{figure}

The drum is filled by allowing 5000 dry particles to fall into it under their own weights 
until a stabilized granular bed is obtained. The number of particles was limited to 5000 
in order to simulate a large number of drum rotations and many runs with different values 
of the system parameters in a reasonable computation time. In all simulations, 
the filling level was $s_f = \frac{2h}{d_c} \simeq 0.56$, where $h$ is the filling height of 
the granular material inside the drum.  We considered three different 
size classes in a range $[d_{min},d_{max}]$ with a size ratio $\alpha = d_{max}/d_{min}$. 
Each size class has the same total volume so that the size distribution is uniform in terms of  
particle volumes. This corresponds to a small number of large particles and a large number of small particles. 
This distribution generally leads to large packing fraction since  small particles optimally fill the pore space 
between the large particles \cite{Voivret2007,Voivret2009}. The mean particle diameter $\langle d \rangle$ 
can also be defined as a function of $\alpha$ by 
\begin{equation}
\langle d \rangle = d_{min} \frac{2\alpha}{1+\alpha} = d_{max}\frac{2}{1+\alpha} 
\label{eq:d}
\end{equation}
 Fig.~\ref{fig:process}(a) shows an example of the initial state in a system with $\alpha=5$ and 
$\langle d \rangle$ = 16 $\mu$m. Fig.~\ref{fig:process}(b) displays the granular flow 
after $50$ rotations. At the beginning, all particles are at rest inside the granulator. 
With drum rotation, a stable flow configuration is reached after nearly four rotations. 

\begin{figure}[tbh]
\centering
\includegraphics[width=0.42\textwidth,clip]{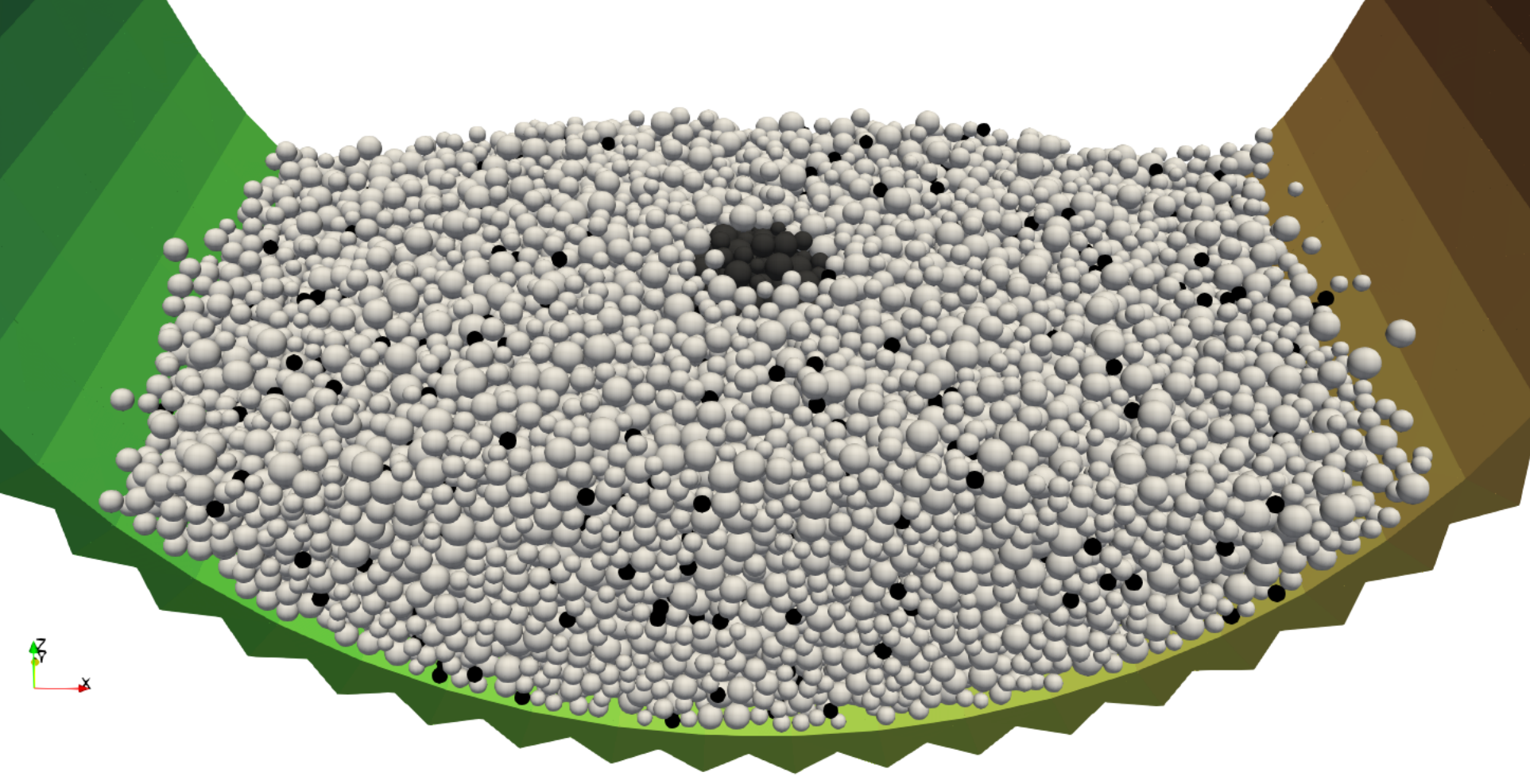}
\caption{Snapshot of the granular bed  showing the distribution of dry particles (in white) and 
wet particles (in black) both those inside the initially defined granule in the center of the bed and 
those randomly distributed throughout the bed.}
\label{fig:sample}
\end{figure}

A number $N_w$ of particles distributed randomly inside the drum can be wetted by 
attributing them a given amount of liquid. These wet particles collide and nucleate 
into small granules during the steady flow of the particles. However, because of the 
relatively low number of particles, it is numerically more efficient to start the simulations 
with a single granule introduced initially in the center of the granular bed. The granulation process 
can then be analyzed by following the particles captured by (accretion) or subtracted from (erosion) 
the granule. Obviously, the coalescence of granules can be investigated only by simulations 
with a much larger number of granules. In this paper, we focus on the first option with granulation 
around a single wet granule. To define the initial wet granule, we place a spherical probe in 
the centre of the granular bed with a radius such that exactly $N_{g0}=100$ particles are inside the probe. 
All these particles are considered to be wet. In addition, we randomly select $N_w - N_{g0} =200$ 
free particles throughout the sample and consider them to be wet. Fig. \ref{fig:sample} displays the 
initial state of the granular bed together with the initial granule and free wet particles. 
The capillary and viscous forces are activated for all wet particles. 

The liquid content of wet particles $w = V_\ell/V_g$, where $V_\ell$ is the amount of the liquid and 
$V_g$ is the particle volume, is assumed to have the same value for all wet particles. 
In all our granulation simulations, we set $w=0.09$, which is sufficient to create wet granules  
in a horizontal rotating drum \cite{Aguado2013}. When two wet particles meet, 
the volume of the liquid bridge is calculated from $V_\ell = w V_g$ transported by 
each of the particles. We also assume that there is no excess liquid so that 
a wet particle can not form a liquid bridge with a dry particle. 

\begin{figure}[tbh]
\centering
\includegraphics[width=0.49\textwidth,clip]{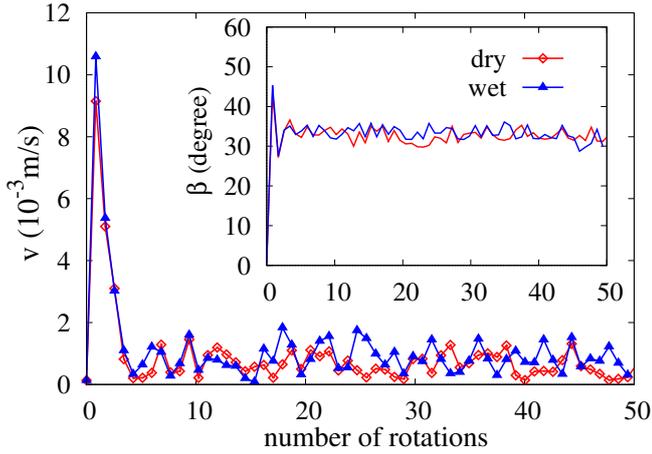}
\caption{The mean velocity $v$ of drum flow for dry particles 
and in the presence of wet particles with $Fr = 0.5$  as a function of the number of rotations of the drum. 
The inset shows the evolution of the free surface inclination angle $\beta$ as a function of the number of rotations.}
\label{fig:velocity}
\end{figure}

We performed a large number of simulations with different values of $d_{min}$ and $d_{max}$ 
in the range $[10,1500]$ $\mu$m, different values of friction coefficient $\mu$ in the range $[0.1,0.9]$ and 
four values of liquid viscosity $\eta$. For each set of values, several independent granular beds 
were generated and subjected to the granulation process. All values of the system parameters used 
in our simulations are listed in Table \ref{tab:parameters}. The choice of most parameter values is 
guided either by numerical efficiency or by reference to the granulation process of iron ores in 
a rotating drum, e.g. for the density of particles, contact angle and filling level. 
However, for data analysis we rely on dimensionless parameters. For rotating drum a relevant 
dimensionless parameter is the Froude number $Fr$ \cite{Yang2008,Mellmann2001,Liu2013}:
\begin{equation}
Fr = \frac{\omega^2 d_c}{2g}. 
\label{eq:fr}
\end{equation} 
The flow regime depends on the value of the Froude number.  
For dry particles, the value $Fr = 0.5$ leads to a flow regime intermediate between rolling and cascading 
regimes \cite{Ennis1996,lefebvre2013}. Since these regimes have been established for 
dry particles, we investigated the effect of the wet particles by comparing the velocity profiles  for 
dry particles, on the one hand, and in the presence of wet particles, on the other hand, 
for $Fr = 0.5$. We observe practically the same velocity profiles in both cases.  
Fig. \ref{fig:velocity} shows the mean particle velocity $v$ and free surface 
inclination $\beta$ for the two simulations with $\alpha=5$. We see that both $v$ and $\beta$  
have the same value in both flows. The only difference is that there are more fluctuations in 
the presence of free wet particles that can be attributed to the higher inhomogeneity of 
the flow in this case.  

\begin{table}
\centering
\caption{Simulation parameters}
\label{tab:parameters}
\begin{tabular}{llll}
\hline
Parameter & Symbol & Value and Unit \\
\hline
Particle diameters & $d$ & [10,1500] $\mu$m \\
Density of particles & $\rho$ & 3500 kg.m$^{-3}$ \\
Size ratios & $\alpha$ & [1,5] \\
Number of particles & $N_p$ & 5000\\
Filling level & $s_f$ & 0.56\\
Friction coefficient & $\mu$ & [0.1,0.9] \\
Normal stiffness & $k_n$ & 100 N/m \\
Tangential stiffness & $k_t$ & 80 N/m \\
Normal damping & $\gamma_n$ & $5.10^{-5}$ Ns/m \\
Tangential damping & $\gamma_t$ & $5.10^{-5}$ Ns/m \\
Surface tension  & $\gamma_s$ & 0.021 N/m \\
Contact angle  & $\theta$ & 4.0 degree \\
Liquid viscosity & $\eta$ & [10;20;40;60] mPa.s \\
Time step & $\delta t$ & 10$^{-7}$ s \\
\hline
\end{tabular}
\end{table}

\section{Parametric study of granulation}
\label{sec:behavior}

In this section, we are interested in the evolution of the granule size, in terms of the 
total number of particles $N_g$ embodied in the granule, as well as the relative contributions of 
the accretion and erosion events. The cumulative accretion is the number $N_g^+$ of free wet particles 
captured by the granule whereas the cumulative erosion $N_g^-$ is the number of wet particles leaving the granule. 
The rates of these events depend on the relative importance of force chains and cohesive stresses 
acting on the granule. The values of the process parameters affect the rates so that the granule may 
grow at different rates. When the rate becomes negative, the granule initially inserted into the granular bed 
will disappear by excess erosion. Unless explicitly stated, the liquid properties are those of water 
($\eta =1$ mPa.s and $\gamma_s = 0.072$ N/m). From the parametric study, 
we will determine the phase diagram of granule growth for polydispersity vs. mean particle size.   

\subsection{Growth, accretion and erosion}

\begin{figure}[tbh]
\centering
\includegraphics[width=0.48\textwidth,clip]{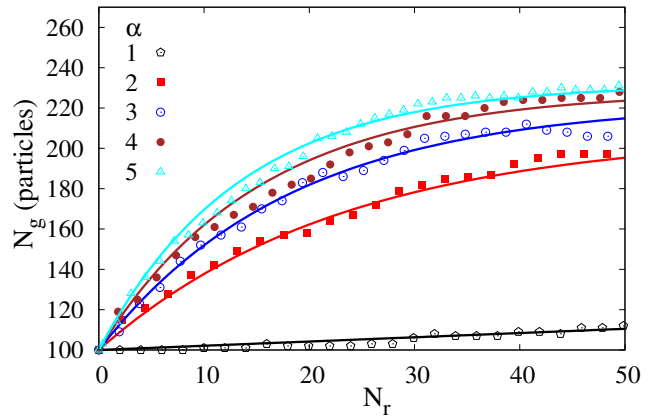}
\caption{Evolution of the granule size $N_g$ (in number of particles) for different values of size ratio $\alpha$ 
and $d_{min}=10$ $\mu$m. The solid lines are exponential fits given by equation (\ref{eqn:ngt}).}
\label{fig:growth_size_ratio}
\end{figure}

\begin{figure}[tbh]
\centering
\includegraphics[width=0.48\textwidth,clip]{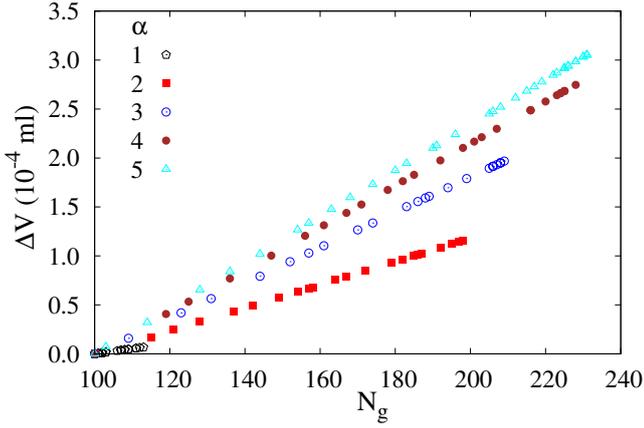}
\caption{Variation $\Delta V$ of the volume of particles inside the granule as a function of their number  $N_g$  
for different values of size ratio $\alpha$ and $d_{min}=10$ $\mu$m.}
\label{fig:N_g_delta_V}
\end{figure}

Figure \ref{fig:growth_size_ratio} shows the evolution of the granule size  
as a function of the number of drum rotations $N_r = \omega t /2\pi$, where $t$ is 
the granulation time,  for different values of the size ratio $\alpha$, $Fr=0.5$ 
and $\mu =0.5$. In these simulations, $d_{min}$ is kept constant and equal to 10 $\mu$m. 
This means that $\alpha$ is increased here by increasing $d_{max}=\alpha d_{min}$. The granule grows  
almost exponentially with $N_r$ at a rate that increases with $\alpha$. In other words, 
the increased size polydispersity enhances granulation. This effect is more spectacular 
when compared to the mono-disperse case ($\alpha=1$) in which the granule growth 
is negligibly small after 50 rotations. As we are interested in this paper in the evolution of captured and 
eroded particles, the granule size $N_g$ is expressed in terms of the number of wet 
particles in the granule. Even for broad size polydispersity, we find that 
the total volume of wet particles belonging to the granule is a linear function of 
their number, as shown in Fig. \ref{fig:N_g_delta_V}, up to a factor that depends on the 
size polydispersity. Hence, the trends of $N_g$ 
investigated below are similar if the granule size is measured in terms of the total volume of 
the granule or the total volume of particles in the granule.         

\begin{figure}[tbh]
\centering
\includegraphics[width=0.45\textwidth,clip]{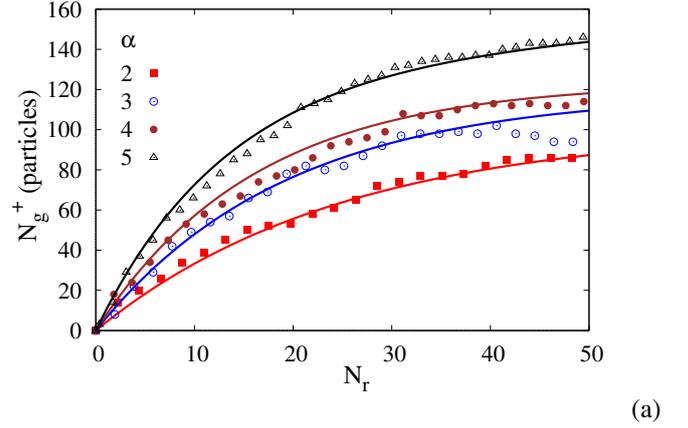}(a)
\includegraphics[width=0.45\textwidth,clip]{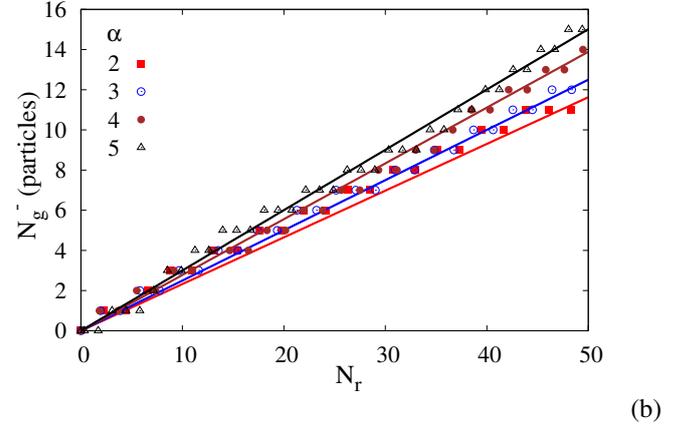}(b)
\caption{Cumulative accretion (a) and cumulative erosion (b) of particles for different values of 
size ratio $\alpha$. The lines are fitting forms given by equations (\ref{eqn:ng-t}) and (\ref{eqn:ng+t}).}
\label{fig:accretion-erosion}
\end{figure}

The cumulative accreted and eroded particles are plotted 
in Fig. \ref{fig:accretion-erosion} as a function of $N_r$ only for poly-disperse samples in which 
we have a significant number of particles captured and eroded. The accretion 
grows exponentially at a negative rate whereas erosion is a linear function. 
We will see below with other values of 
material parameters that the erosion can grow in a non-linear way and faster than accretion so that the 
linear evolution observed in Fig. \ref{fig:accretion-erosion} may be considered as 
a first-order effect in the limit of low erosion rates.  
Fig. \ref{fig:accretion-erosion} also shows that both accretion and erosion increase with $\alpha$ to an extent that 
is higher for accretion than for erosion.

The exponential increase of the number of accreted particles $N_g^+$ is a consequence of the 
decreasing number of available free wet particles in the granular bed while they are 
captured by the granule. The number of free wet particles is given by $N_w - N_g$, 
where $N_w$ is the total number of wet particles including those belonging to the granule. 
This is equivalent to the decrease of the available liquid for granulation as the granules grow. 
Hence, in the steady flow, we may assume that  the variation $\Delta N_g^+$ 
of the captured particles is proportional to the current number $N_w - N_g$ of wet particles 
and to the elapsed time $\Delta t$ or angular rotation $\Delta N_r = \omega \Delta t$:
\begin{equation} 
\Delta N_g^+ = k^+ \frac{N_w - N_g}{N_w} \Delta N_r, 
\label{eqn:ng+}
\end{equation}
where $k^+$ is the relative accretion rate. 
As to the number $N_g^-$ of eroded particles, we assume that  its rate $k^-$ is constant : 
\begin{equation}
\Delta N_g^- = k^- \Delta N_r, 
\label{eqn:ng-}
\end{equation}
These equations are consistent with our numerical data shown in Fig. \ref{fig:accretion-erosion} although we expected 
the number of eroded particles to be proportional to the number of particles at 
the surface of the granule. This effect may reflect the fact that the average curvature of the 
granule surface declines as its size increases so that the particles lying at the surface of the granule 
are more strongly attached to the granule and/or less subjected to the eroding action of 
granular flow. This effect may counter-balance the increase of the granule surface area. 
However, for much larger granules in number of primary particles this effect may disappear.      

With the above assumptions, the rate equation for the granule size is 
simply $\Delta N_g = \Delta N_g^+ - \Delta N_g^-= (- k^+ N_g/N_w + k^+  - k^-) \Delta N_r$, which 
leads to a simple differential equation:
\begin{equation}
\frac{d N_g}{dN_r} = k^+ \left(1-N_g/N_w\right)   - k^-
\label{eqn:ngdt}
\end{equation}
with the following solution: 
\begin{equation} 
N_g (N_r) = N_{g0} + \{N_w( 1 - k^-/k^+) - N_{g0}\} (1 - e^{\frac{-k^+}{N_w} N_r})
\label{eqn:ngt}
\end{equation} 
This model predicts an exponential growth and an asymptotic  granule size $N_g (t \to \infty)= N_w( 1- k^-/k^+)$, 
as observed in our simulations. The steady granulation 
state corresponds to the condition  $\Delta N_g^+=\Delta N_g^-$. For $k^- \ll k^+$, 
the final granule embodies nearly all wet particles: $N_g \simeq N_w$. On the other 
hand, the granule disappears if $k^- >  k^+ $.  

\begin{figure}[tbh]
\centering
\includegraphics[width=0.45\textwidth,clip]{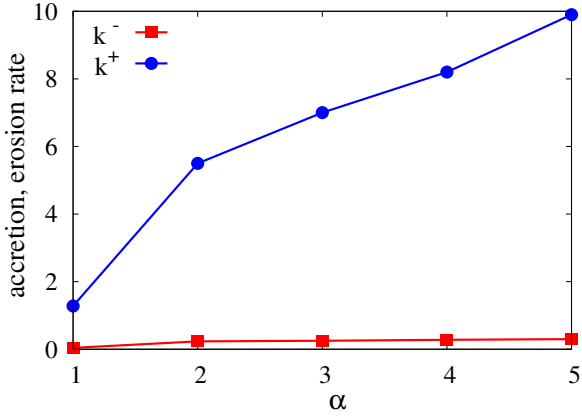}
\caption{Fitted values of erosion rate $k^-$ and accretion rate $k^+$ as a function 
of polydispersity parameter $\alpha$.}
\label{fig:alpha-k}
\end{figure}

The evolution of $N_g^+$ and $N_g^-$ as a function of drum rotation can be obtained 
from equations (\ref{eqn:ngt}), (\ref{eqn:ng+}) and (\ref{eqn:ng-}). We get 
\begin{equation} 
N_g^- = k^- N_r
\label{eqn:ng-t}
\end{equation} 
and 
\begin{equation} 
N_g^+ =  \{N_w(1  - k^-/k^+) - N_{g0}\} (1 - e^{\frac{-k^+}{N_w} N_r}) + k^- N_r
\label{eqn:ng+t}
\end{equation}
In view of the present model, the influence of particle size ratio $\alpha$ observed in Figs. \ref{fig:growth_size_ratio} 
and \ref{fig:accretion-erosion} can be interpreted in terms of the accretion and erosion rates. 
Fig.\ref{fig:alpha-k} shows the fitted values of $k^-$ and $k^+$  
as a function of $\alpha$. 
The increase of accretion rate $k^+$ with $\alpha$ is rather counter-intuitive since the cohesive strength 
is inversely proportional to the mean particle size, which increases here with $\alpha$. This means that 
the higher polydispersity, allowing for a better filling of the pore space and thus higher density of the granule, 
over-compensates the decrease of the cohesive stress. But the latter explains 
the increase of the erosion rate $k^-$, which is quite small compared to $k^+$.  

\subsection{Effects of material parameters}
\label{subsec:material}

\begin{figure}[tbh]
\centering
\includegraphics[width=0.50\textwidth,clip]{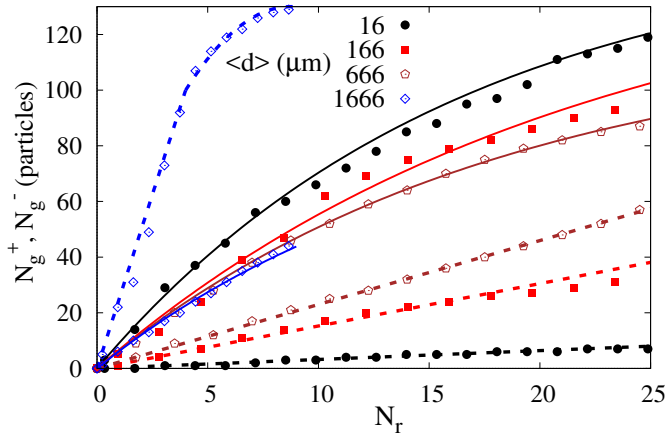}
\caption{Evolution of the cumulative number of wet particles for accretion 
(solid lines) and erosion (dashed lines) for four different values of the mean particle 
diameter $\langle d \rangle$ and size ratio $\alpha=5$, as a function of the 
number of drum rotations.}
\label{fig:size_acc_ero}
\end{figure}

We now consider the effect of the mean particle size $\langle d \rangle$, which 
directly controls the cohesive stress of wet particles. Fig. \ref{fig:size_acc_ero} 
displays the cumulative number of particles for both accretion and erosion in the case 
$\alpha=5$, $\mu=0.5$. Note that, in these simulations, the higher values of $\langle d \rangle$ 
imply higher values of both $d_{min}$ and $d_{max}$. But, according to equation (\ref{eq:d}), $d_{max}$ 
increases faster  than $d_{min}$. Fig. \ref{fig:size_acc_ero} shows that accretion 
$N_g^+$ increases as an exponential function of the number of drum rotations whereas erosion 
$N_g^-$ is quasi-linear. As expected, since the cohesive stress declines, accretion decreases 
and erosion increases with increasing  $\langle d \rangle$. For $\langle d \rangle = 1666$ $\mu$m, 
erosion is high enough to cancel the effect of accretion, and thus the granule disappears after 8 rotations. 
Here, the cumulative erosion does not grow linearly with time and for this reason the erosion rate is not 
constant. For the linear part of erosion, we have $k^- \simeq 25$ whereas $k^+ \simeq 6$. Clearly, $k^-$ is 
strongly dependent on the cohesive stress of the granule, which declines in inverse proportion 
to the wet particle mean size. The cohesive stress of wet particles affects, albeit to a lesser extent, 
the accretion rate $k^+$.    

\begin{figure}
\centering
\includegraphics[width=0.51\textwidth,clip]{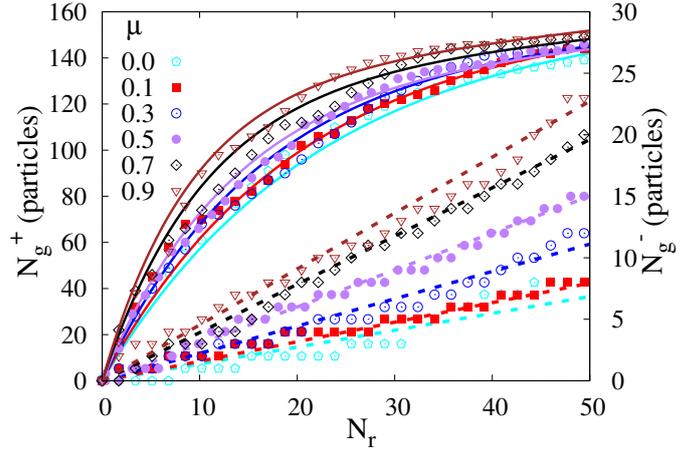}
\caption{Cumulative accretion (solid lines) and erosion (dashed lines) of wet particles for five different values of 
the friction coefficient $\mu$, size ratio $\alpha=5$ and $d_{min}=10$ $\mu$m, 
as a function of the number of drum rotations. The lines are fitting forms given by equations (\ref{eqn:ng-t}) and (\ref{eqn:ng+t}).}
\label{fig:friction_acc_ero}
\end{figure}

\begin{figure}
\centering
\includegraphics[width=0.45\textwidth,clip]{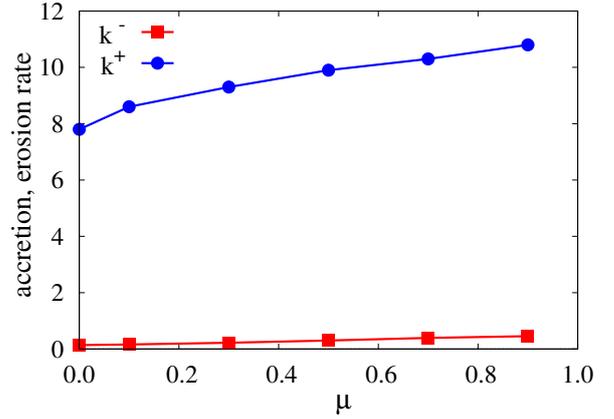}
\caption{Fitted values of erosion rate $k^-$ and accretion rate $k^+$ as a function of friction coefficient $\mu$ 
for size ratio $\alpha=5$.}
\label{fig:mu-k}
\end{figure}

The friction coefficient $\mu$ between particles is a major parameter for 
granular flows. Its joint effect with cohesive forces can thus influence the granulation process.   
Fig. \ref{fig:friction_acc_ero} shows cumulative accretion and erosion for an increasing value of 
$\mu$ and exponential fits. 
Fig. \ref{fig:mu-k} displays $k^-$ and $k^+$ as a function of $\mu$. 
We observe a slight increase of both rates. The values of $N_g^+$ and $N_g^-$ 
after 50 rotations show that   
accretion increases slightly with $\mu$, which may be understood 
as a consequence of enhanced capturing of free wet particles when they touch the granule. 
Surprisingly, however, $N_g^-$ increases to a larger extent with $\mu$ so that 
the granule growth is slower at larger values of $\mu$. As erosion is a consequence 
of interactions between dry particles and 
boundary wet particles of the granule, the increase of erosion with $\mu$ 
may be understood as an increase of shear forces acting on the boundary particles 
by shear flow of dry particles. This suggests that the granulation of rough particles 
(with higher friction coefficient) is less efficient than rounded particles and 
it should consume more energy.

\begin{figure}
\centering
\includegraphics[width=0.51\textwidth,clip]{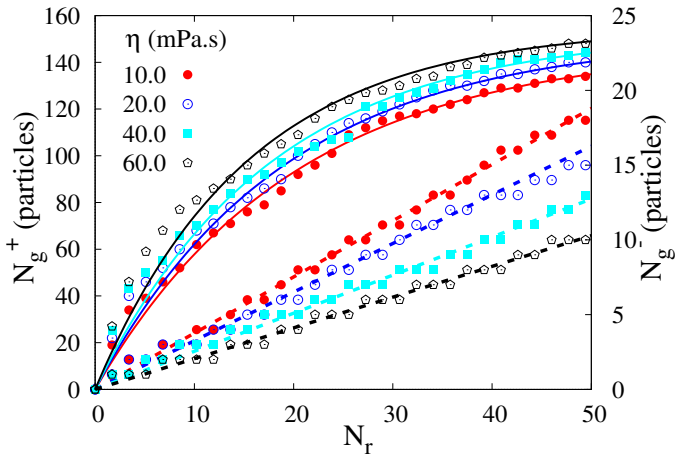}
\caption{Cumulative accretion (solid lines) and erosion (dashed lines) for four different values 
of the liquid viscosity $\eta$ for size ratio $\alpha=5$ and $d_{min}=10$ $\mu$m as a function 
of the number of drum rotations. The solid and dashed lines are exponential and linear 
 fits to the data points, respectively.}
\label{fig:liquid_acc_ero}
\end{figure}

\begin{figure}
\centering
\includegraphics[width=0.45\textwidth,clip]{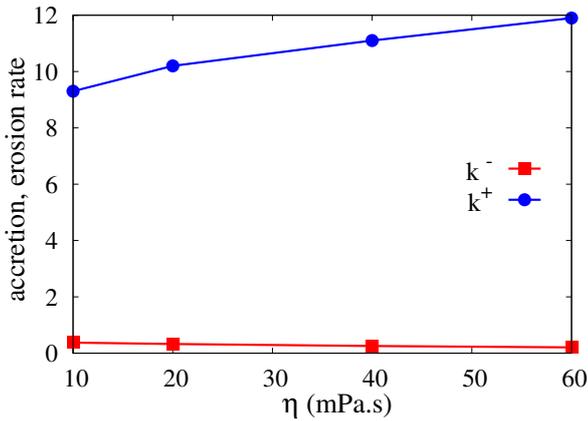}
\caption{Erosion rate $k^-$ and accretion rate $k^+$  
as a function of liquid viscosity $\eta$ for size ratio $\alpha=5$ and $d_{min}=10$ $\mu$m.}
\label{fig:k-eta}
\end{figure}

This brings us to the effect of liquid viscosity expressed as a lubrication force 
between wet particles. We performed several granulation simulations for different values of 
$\eta$ in the range $[10,60]$ mPa.s and with same value of the liquid-vapor surface tension 
$\gamma_s$=21 mN/m. Fig. \ref{fig:liquid_acc_ero} shows the evolution of 
cumulative accretion and erosion for $\alpha=5$. Again, we observe the exponential increase of $N_g^+$ vs. 
the nearly linear increase of $N_g^-$ with the number of drum rotations. 
The accretion increases slightly with $\eta$ whereas the erosion declines. 
The decrease of $N_g^-$ is exactly the opposite effect of friction coefficient in 
Fig. \ref{fig:friction_acc_ero} which causes an increase of $N_g^-$. This means that  
lubrication forces tend to reduce the shearing effect of the flow on the granule, leading to 
smaller erosion. On the other hand, the increase of accretion can be attributed to 
viscous dissipation that can enhance the capture of free wet particles by the 
granule.  Fig. \ref{fig:k-eta} shows $k^-$ and $k^+$ 
as a function of $\eta$. The variations of the rates with $\eta$ are small but their 
opposite effects tend to enhance granule growth.     

\subsection{Phase diagram}
\label{subsec:phase}

\begin{figure}
\centering
\includegraphics[width=0.49\textwidth,clip]{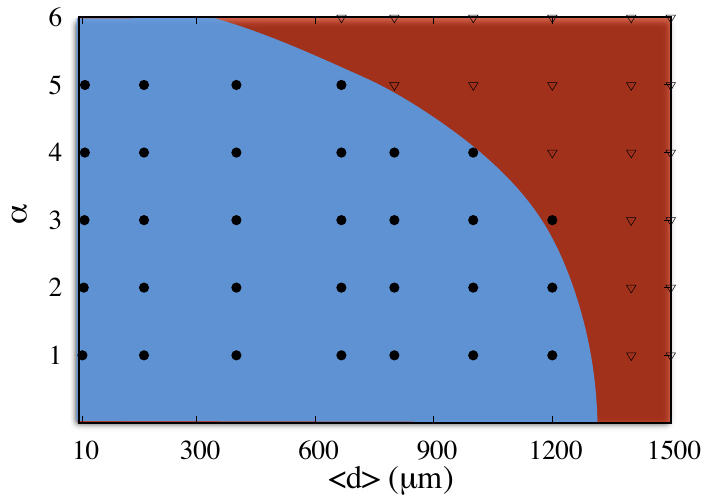}
\caption{Phase diagram of granule growth in the parametric space of $\langle d \rangle$ 
vs. $\alpha$ for $\eta$ = 1 mPa.s and $\mu=0.5$. The granule grows only in the light blue region and 
disappears otherwise.}
\label{fig:alpha_d}
\end{figure}

In the simulations reported in this paper, we have a single granule of size $N_{g0}$ 
that can increase or decrease in size depending on the values of various material parameters. 
In the last section, we analyzed the influence of several parameters on the accretion and erosion rates. 
The most crucial issue, however, in this single-granule problem is the ranges of the values of 
those parameters for which the granule will survive and grow, i.e. the phase diagram of granulation. 
We consider here only the phase diagram in the parameter space of the mean particle size 
$\langle d \rangle$ vs. size ratio $\alpha$. In these simulations, we set $\gamma_s$=72 mN/m and $\mu=0.5$. 
   
Figure \ref{fig:alpha_d} displays all the simulated points as a grid in the parameter space 
$[\langle d \rangle,\alpha]$ and an approximate frontier between the range of values for which 
the granule survives, i.e. granulation is possible, and the values for which no granule can grow. 
The liquid viscosity is set to $\eta=1$ mPa.s. We see that for  $\langle d \rangle > 1200$ $\mu$m 
no granulation occurs even for larger values of $\alpha$ (larger polydispersity). 
For smaller values of $\langle d \rangle$, the range of the values of $\alpha$ for granulation increases. 
Note that the limit on the value of $\alpha$ is related to the fact that for given $\langle d \rangle$, 
the increase of $\alpha$ requires the increase of $d_{max}$ and decrease of $d_{min}$. 
The increase of $d_{max}$ compensates to some extent the effect of decreasing 
$d_{min}$ on the cohesive stress.      


\section{Conclusions}
\label{sec:conclude}

In this paper, we used a 3D molecular dynamics algorithm with a capillary cohesion 
law enhanced by the viscous effect of the binding liquid in order to investigate the 
granulation process of granular materials in a horizontal drum rotating about its axis.  
The system was numerically prepared by pouring solid particles into a rigid drum 
composed of polyhedral elements. We considered in detail the evolution of a 
single granule introduced from the beginning into the granular bed for a broad range of 
parameter values. The liquid was assumed to be transported by the particles and 
its amount was defined by the number of wet free particles randomly distributed 
inside the granular bed.  

We showed that the granule grows almost exponentially with time 
(or the number of drum rotations) as a result of a gradual capture of 
free wet particles by the granule. The accretion of free wet particles 
is nearly always an exponential function of time whereas the number of eroded 
particles grows linearly with time. A simple model based on constant accretion and 
erosion  rates was introduced, predicting the observed exponential increase 
of granule size leveling off to a constant value at long times. We investigated 
the effects of size ratio, mean particle size, friction coefficient and liquid viscosity 
on accretion and erosion of particles. Both accretion and erosion increase when 
the size ratio or friction coefficient are increased. Accretion declines whereas 
erosion increases when the mean particle size is increased. Accretion increases 
whereas erosion declines when liquid viscosity is increased. We determined the 
phase diagram of granulation by varying systematically the size ratio and mean particle size. 
This will be extended to other parameters in the future. 

Our results are based on a simple system with a relatively small number of spherical particles. 
We considered the evolution of a single granule and  this allowed us to perform long-time 
simulations for a range of values of material parameters. Clearly, these simulations can be extended 
to obtain the combined effects of parameters in phase-space diagrams. 
The effects of process parameters such as filling rate, rotation speed and 
wetting procedure can be studied, too. Further data analysis is also necessary in order to 
investigate the granule consolidation and the problem of nucleation from free wet particles 
when no granule is initially introduced. Despite its higher computational cost, 
a larger number of particles may allow for the simulation of multi-granule systems 
and coalescence phenomena.      

\section*{Acknowledgments}

We gratefully acknowledge financial support by the Ministry of Education and Training in Vietnam and 
Campus France. This work was initiated in collaboration with ArcelorMittal.

\bibliographystyle{elsarticle-num}

\end{document}